\documentclass[fleqn,usenatbib]{mnras}

\usepackage{newtxtext,newtxmath}

\usepackage[T1]{fontenc}
\usepackage{ae,aecompl}

\usepackage{graphicx}	
\usepackage{amsmath}	
\usepackage{amssymb}	

\newcommand{\tic}{TIC-238855958}	
\newcommand{\tess}{\textit{TESS}}	%
\newcommand{\wasp}{\textit{WASP}}	%
\newcommand{\rsun}{\ensuremath{R_{\sun}}} %
\newcommand{\teff}{\ensuremath{T_{\rm eff}}} %

\usepackage[T1]{fontenc}
\usepackage{ae,aecompl}

\usepackage{newtxtext,newtxmath}

\title[Photometric recovery of a TESS single-transit candidate]{NGTS and WASP photometric recovery of a single-transit candidate from {\it TESS}}

\author[0000-0002-4259-0155]{
\parbox{\textwidth}{Samuel Gill$^{1,2}$, 
Daniel Bayliss$^{1,2}$,
Benjamin F. Cooke$^{1,2}$,
Peter J. Wheatley$^{1,2}$,
Louise~D.~Nielsen$^{3}$, 
Monika~Lendl$^{3,4}$,
James McCormac$^{1,2}$,
Edward~M.~Bryant$^{1,2}$,
Jack~S.~Acton$^5$,
David R. Anderson$^{1,2}$,
Claudia~Belardi$^5$,
Fran\c{c}ois~Bouchy$^{3}$,
Matthew~R.~Burleigh$^5$,
Andrew ~Collier~Cameron$^6$,
Sarah~L.~Casewell$^5$,
Michael~R.~Goad$^5$,
Maximilian~N.~G{\"u}nther,$^{7,8}$
Coel Hellier$^9$,
James A. G. Jackman$^{1,2}$,
James~S.~Jenkins$^{10,11}$,
Maximiliano~Moyano$^{12}$,
Don Pollacco,$^{1,2}$,
Liam~Raynard$^5$,
Alexis~M.~S.~Smith$^{13}$,
Rosanna~H.~Tilbrook$^5$,
Oliver~Turner$^{3}$, 
St\'{e}phane~Udry$^{3}$, 
Richard G. West$^{1,2}$,
 }\\
$^{1}$ Department of Physics, University of Warwick, Gibbet Hill Road, Coventry CV4 7AL, UK \\
$^{2}$ Centre for Exoplanets and Habitability, University of Warwick, Gibbet Hill Road, Coventry CV4 7AL, UK \\ 
$^{3}$Observatoire de Gen{\`e}ve, Universit{\'e} de Gen{\`e}ve, 51 Ch. des Maillettes, 1290 Sauverny, Switzerland \\
$^4$ Space Research Institute, Austrian Academy of Sciences, Schmiedlstr. 6, 8042 Graz, Austria\\
$^5$ School of Physics and Astronomy, University of Leicester, Leicester LE1 7RH, UK\\
$^6$ Centre for Exoplanet Science, SUPA, School of Physics and Astronomy, University of St Andrews, St Andrews KY16 9SS, UK\\
$^{7}$Department of Physics, and Kavli Institute for Astrophysics and Space Research, Massachusetts Institute of Technology\\
Cambridge, MA 02139, USA \\
$^{8}$Juan Carlos Torres Fellow\\
$^9$ Astrophysics Group, Keele University, Staffordshire, ST5 5BG, UK\\
$^{10}$ Departamento de Astronom\'ia, Universidad de Chile, Camino El Observatorio 1515, Las Condes, Santiago, Chile\\
$^{11}$ Centro de Astrof\'isica y Tecnolog\'ias Afines (CATA), Casilla 36-D, Santiago, Chile\\
$^{12}$Instituto de Astronom\'ia, Universidad Cat\'olica del Norte, Angamos 0610, 1270709, Antofagasta, Chile. \\
$^{13}$Institute of Planetary Research, German Aerospace Center, Rutherfordstrasse 2, 12489 Berlin, Germany\\
}

\date{Last updated 2015 May 22; in original form 2013 September 5}

\pubyear{2015}

\begin{document}

\label{firstpage}
\pagerange{\pageref{firstpage}--\pageref{lastpage}}
\maketitle

\begin{abstract}
The Transiting Exoplanet Survey Satellite (\tess) produces a large number of single-transit event candidates, since the mission monitors most stars for only $\sim$27\,days.  Such candidates correspond to long-period planets or eclipsing binaries.  Using the \tess\ Sector 1 full-frame images, we identified a 7750\,ppm single-transit event with a duration of 7\,hours around the moderately evolved F-dwarf star \tic\ (Tmag=10.23, \teff=6280$\pm{85}$\,K).  Using archival WASP photometry we constrained the true orbital period to one of three possible values.  We detected a subsequent transit-event with NGTS, which revealed the orbital period to be 38.20\,d.  Radial velocity measurements from the CORALIE Spectrograph show the secondary object has a mass of $M_2$= $0.148\pm{0.003}$\,M$_{\odot}$, indicating this system is an F-M eclipsing binary.  The radius of the M-dwarf companion is $R_2$ = $0.171\pm{0.003}$\,R$_{\odot}$, making this one of the most well characterised stars in this mass regime.  We find that its radius is 2.3-$\sigma$ lower than expected from stellar evolution models. 
\end{abstract}




\begingroup
\let\clearpage\relax
\endgroup


\section{Introduction}\label{sec:intro}
The Transiting Exoplanet Survey Satellite \citep[\tess,][]{2015JATIS...1a4003R} has successfully completed its Year 1 survey of the southern ecliptic hemisphere, yielding over 1000 \tess\ Objects of Interest(TOIs).  Already many of these systems have been confirmed as bona fide transiting exoplanets \citep[e.g. ][]{2018ApJ...868L..39H,2019ApJ...881L..19V,2019A&A...623A.100N,2019NatAs.tmp..409G}. The majority of the TOIs are short period systems, with the mean orbital period of these at 7.89\,days.  This is due to the geometric probability of a transit transit being inversely proportional to the planet's semi-major axis, and the fact that \tess\ only monitors most stars for a single 27\,day Sector.  In order to discover longer period systems in the \tess\ data we need to follow-up and characterise the systems that only present a single-transit event in the \tess\ light curves.  It is clear that there will be a large number of these single-transit candidates in the \tess\ data \citep{2018A&A...619A.175C, 2019AJ....157...84V}, and these longer period systems are scientifically valuable.  For low mass host stars (late K and early M) it will allow us to probe planets in the habitable zones.  It is also important to study longer period eclipsing binaries as it allows us to study the mass-radius relationship for low mass stars without the complicating effects of high stellar irradiation and strong tidal interactions.

With these longer-period \tess\ systems in mind, we have begun a program within the ambit of the NGTS project \citep{2018MNRAS.475.4476W} to find and characterise single-transit event candidates.  This paper reports the first result from this program in which we determine photometrically the orbital period of the \tess\ single-transit candidate \tic.  We also use spectroscopy to
measure the mass of the secondary companion, revealing the system to be a long period F-M binary.  

In Sect.\,\ref{sec:single}, we outline our single-transit event search of \tess\ data, which led to the discovery of \tic.
We describe our search of archival WASP photometry
in Sect.\,\ref{sec:precovery}, and in Sect.\,\ref{sec:observations} we describe our use of the NGTS facility to recover a subsequent transit of \tic.  We analyse all the data and derive system parameters in Sect.\,\ref{sec:analysis}, and finish with our conclusions in Sect.\,\ref{sec:conclusion}.


\section{Single-Transit Event Detection}\label{sec:single}
We conducted a systematic search of \tess\ light curves for single transit events.  We began by downloading the difference-imaging full-frame light curves produced using the pipeline from \citet{2018AJ....156..132O}.  These light curves are publicly available on Filtergraph\footnote{https://filtergraph.com/tess\_ffi/sector-01}. We downloaded all light curves with a contamination ratio $<5$. Before beginning our search, we flattened each light curve using the {\sc Lightkurve} tool, part of the publicly-available Kepler/K2 community tools \citep{2018ascl.soft12013L}. These flattened light curves were then searched systematically using the method set out in \citet{2016MNRAS.457.2273O}, searching for single-transit events up to 24\,h in duration.  We detected on order of 1000 high signal-to-noise-ratio events per sector. 
We vet these candidate events further by checking for known systematics, known planets or eclipsing binaries, checking Gaia DR2 \citep{2018A&A...616A...1G} for blends and analysing the \textit{TESS} full-frame images (FFIs) for asteroids and other external influences.

\begin{figure}
    \centering
    \includegraphics[scale=0.6]{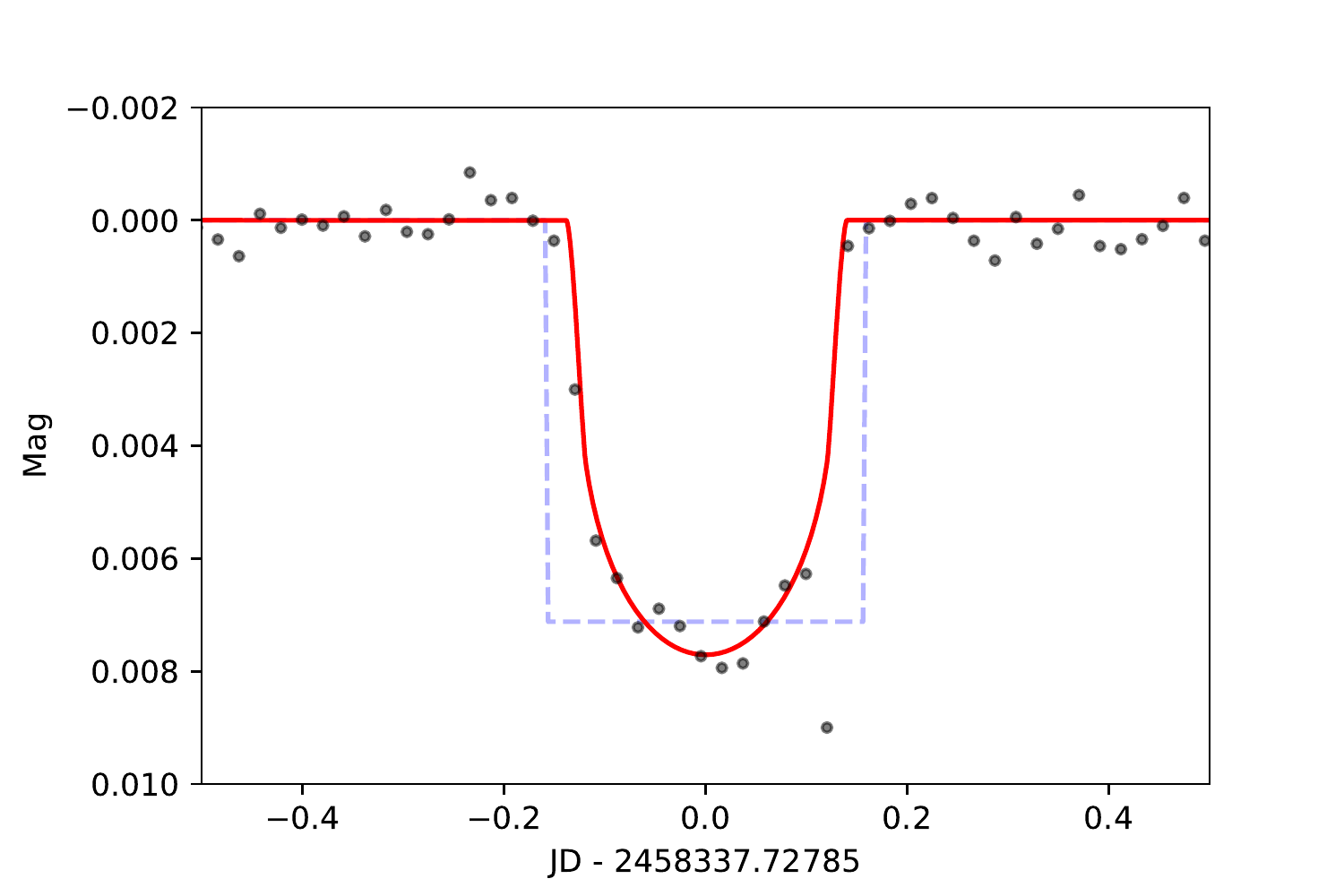}
    \caption{Difference imaging TESS light curve for TIC-238855958 (black) with best-fitting global model (red) and box used to detect the single-transit event (blue-dashed). }
    \label{fig:tess}
\end{figure}

Using this process, \tic\ was identified as a strong single-transit candidate in our search of the \tess\ Sector 1 data.  \tic\ is a T=10.23 magnitude star located at RA=342.750156$^\circ$ and Dec=-67.51508$^\circ$.  From the TESS Input Catalogue 8 \citep{2018AJ....156..102S}, \tic\ is a \teff\ =6200\,K F dwarf with a radius of 2.26\rsun.  \tic\ does not appear in subsequent \tess\ Sectors, so there was no possibility of further transits in the \tess\ data.  We show the flattened difference imaging full-frame light curve for \tic\ in Fig.\,\ref{fig:tess}.  The single transit event has a depth of 7750\,ppm and a duration of 7 hours.  Excluding the transit feature, the light curve of \tic\ shows a RMS of 460 ppm, so the transit feature is clearly significant.  No other stars around \tic\ show a similar transit feature at this epoch, helping rule out a spacecraft systematic.  We see no evidence for any asteroid or other irregularity (including centroid offsets) in the full-frame pixel data that could be responsible for the single-transit event.

\section{Transit precovery with WASP}\label{sec:precovery}

For each single-transit candidate we identify in \tess\ data using the method set out in Section~\ref{sec:single}, we cross-match the star with archival data from the Wide-Angle Search for Planets \citep[WASP; ][]{2006PASP..118.1407P}.  WASP operates two survey instruments: one at the South African Astronomical Observatory (SAAO), South Africa, and another at the Observatorio del Roque de los Muchachos, La Palma. \tic\ was observed for 3 consecutive observing seasons from 2010 to 2012 (1SWASPJ225059.97$-$673054.2) from the south station (27,223 observations in total).

In order to search the archival photometric data for evidence of transit events, we use a template-matching algorithm.  We first construct a best-fitting \tess\ template. Although we used the difference-imaging light curve of \citet{2018AJ....156..132O} for the detection (see Section~\ref{sec:single}), we found that for \tic\ the light curve produced using the \textsc{eleanor} pipeline \citep{2019PASP..131i4502F} was of slightly higher photometric quality. We thus adopted the \textsc{eleanor} light curve for our template matching.  The \textsc{eleanor} light curve of \tic\ was modelled using a Bayesian sampler provided through the python package, {\textsc{emcee}}, \citep{2013PASP..125..306F} using the transit model described in Sect. \ref{Sect:joint}. We fixed the orbital period to 30 days and fitted only the transit epoch, $T_0$, the scaled orbital separation, $R_1 / a$, the ratio of radii, $k=R_2/R_1$, the impact parameter, $b$, and the photometric zero-point, $zp$. Limb-darkening parameters were fixed and interpolated using effective surface temperature ($T_{\rm eff}$) from TESS Input Catalogue 8 assuming solar surface metalicity ([Fe/H]) and surface gravity ($\log g$). We ran 50 Markov chains for 10,000 draws and found best fitting parameters of $R_1 / a = 0.083$, $k=0.080$ and $b=0.12$.
This transit template was then used as a matched filter, fitting it to the WASP photometry at each time point in the dataset and recording the  $\chi^2$ statistic to quantify the goodness-of-fit as a function of transit mid-time.  We first calculate the weighted mean ($w_m$) of the full WASP dataset 
and calculate $\chi^2_{\rm ref} = \sum_i (m_i - w_m)^2 / \sigma_i^2$, where $\sigma_i$ is the magnitude error attributed to each data point. The \tess\ template was centred at each point in the \wasp\ light curve and calculated $\Delta \chi^2 = \chi^2 - \chi^2_{\rm ref}$, where $\chi^2 = \sum_i (m_i - t_i)^2 / \sigma_i^2$ and $t_i$ is the centred \tess\ template at each point in the WASP light curve. Times where the template is well-matched to the data correspond to peaks in $\Delta\log \mathcal{L} = -\Delta\chi^2/2$.
Validating these peaks required sufficient thresholding to ensure relatively small peaks in $\Delta\log\mathcal{L}$ caused by white and red photometric noise were not mistaken for transit-like events. We use an empirically determined threshold of $\Delta\log\mathcal{L} > 35$ as our minimum threshold for selecting single-transit-event candidates. 

\begin{figure*}
    \centering
    \includegraphics[scale=0.5]{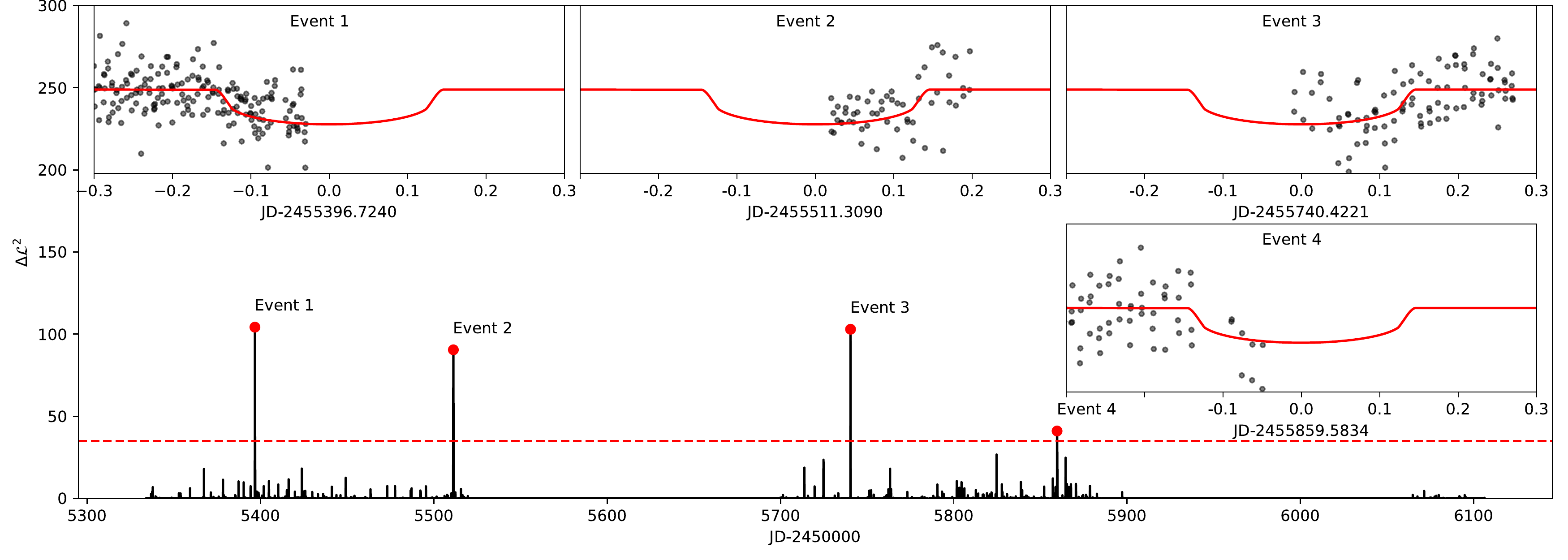}
    \caption{Results of template matching in archival WASP data for \tic. $\Delta \chi^2$ is plotted as a function of time with peaks labelled with red circles when they meet the threshold level (red dashed). The WASP light curve for each matching event is plotted as an inset panel, with the \tess\ template overlaid in red.}
    \label{fig:wasp}
\end{figure*}

For \tic, four significant peaks were identified in $\Delta\log\mathcal{L} > 35$ corresponding to transit-like events in WASP photometry (events 1-4 ordered chronologically; see Fig. \ref{fig:wasp}). Events 1-3 are likely real due to the number of in-transit data points and how well-matched the photometry is to the template. Event 4, which is only just above our threshold, contains only 7 in-transit data points, is likely spurious, and so was excluded. Indeed, the orbital ephemerides from global modelling (Section \ref{Sect:joint}) confirm that only three nights of WASP observations have in-transit data (events 1-3). The maximum orbital period is the smallest elapsed time between transit events, which is 114.585\,d between events 1 and 2. Using the \wasp\ and \tess\ photometry, we are able to determine that the orbital period of \tic\ must be one of three possible orbital periods: $P_{\rm orb} \rm [d] = \{ 114.585, 57.293, 38.195 \}$.  

\section{NGTS TRANSIT DETECTION}\label{sec:observations}

In order to test the set of three possible periods found using the WASP data in Sect.\,\ref{sec:precovery}, we used the Next Generation Transit Survey \citep[NGTS; ][]{2018MNRAS.475.4476W} telescopes located at the ESO Paranal Observatory in Chile.  NGTS was designed for very high precision time-series photometry of stars, and thus is the perfect instrument to use for photometric follow-up of \tess\ single-transit candidates.  The NGTS telescopes are robotic, so it is straightforward to schedule observations based on the possible periods of a given single-transit candidate.  Each NGTS telescope has a field of view of 8 square degrees, providing sufficient reference stars for even the brightest \tess\ candidates.  The telescopes have apertures of 20\,cm and observe at a bandpass of 520-890\,nm.  Full details of the NGTS telescopes and cameras can be found in \citet{2018MNRAS.475.4476W}.

\begin{figure}
    \centering
    \includegraphics[scale=0.6]{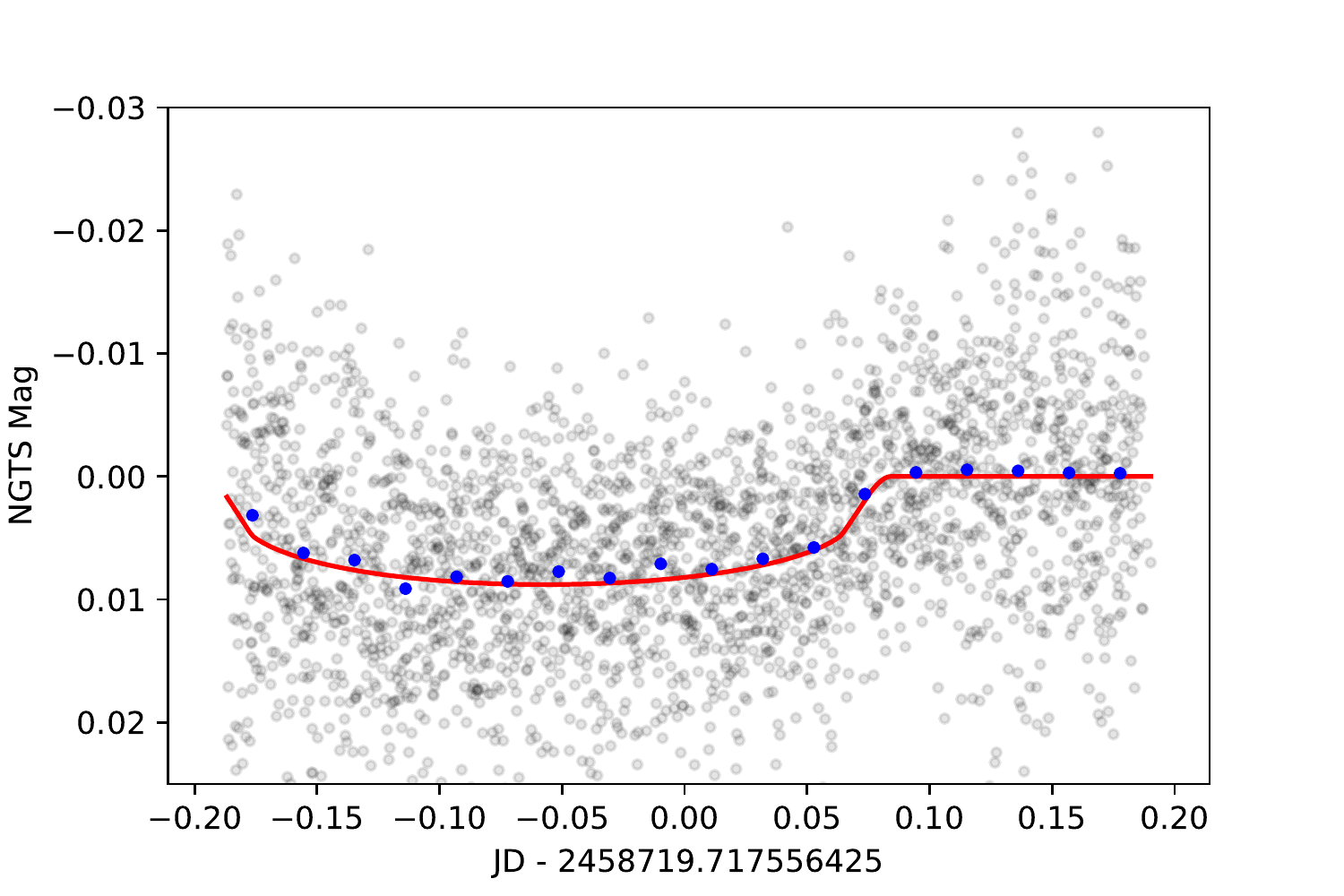}
    \caption{NGTS photometry of \tic\ on the night of 2019 Aug 23. We plot the 30-minute bins (blue) along with the best-fitting global model (red).}
    \label{fig:ngts}
\end{figure}

We scheduled a single NGTS telescope to observe \tic\ on the night of 2019 Aug 23 in order to cover the next visible transit event assuming a 38\,day orbital period.  We observed \tic\ for 9\,hours under photometric conditions with airmass $<$2.  In total we obtain 2\,512 observations, each with exposure times of 10\,s.  Data were reduced on-site using standard aperture photometry routines.  The final light curve is presented in Fig.\,\ref{fig:ngts}, and shows a robust detection of the transit-event for \tic\ that is consistent with the single event seen in the \tess\ data in Sect.\,\ref{sec:single} and the partial events from the WASP data set out in Sect.\,\ref{sec:precovery}.  This confirmed that the only possible orbital period for \tic\ was the 38.195\,day solution.  

\section{Spectroscopic Observations}\label{sec:spec}

In addition to the photometric follow-up of our \tess\ single-transit event candidates set out in Sect.\,\ref{sec:observations}, we also have an ongoing campaign to monitor these candidates spectroscopically using CORALIE - a fiber-fed \'{e}chelle spectrograph installed on the 1.2-m Leonard Euler telescope at the ESO La Silla Observatory \citep{2001A&A...379..279Q,2008ApJ...675L.113W}.  Candidates are vetted with a single observation to check for interlopers such as double lined eclipsing binary systems.  Further observations are then taken spaced appropriately in time in order to determine the mass of the secondary companion.

Following the successful recovery of the orbital period of \tic\ using NGTS, we took eight spectroscopic observations of \tic\ with CORALIE using an exposure time of $t_{\rm exp} = 600$\,s.  The spectra were reduced with the CORALIE standard reduction pipeline, and radial velocity measurements were obtained using standard cross-correlation techniques using numerical masks. We found a high amplitude radial velocity signal of K=7.01$\pm{0.05}$\,km\,s$^{-1}$ which was in phase with the photometric observation - see Fig.\,\ref{fig:Figure_d}.  This amplitude indicated that the companion to \tic\ was in fact stellar in nature, and was also on a moderately eccentric orbit.  We used these radial velocity measurement in our global modelling set out in Sect.\,\ref{Sect:joint}.  We compared the radial velocities to the bisector spans and found no evidence of correlation.

\section{Analysis}\label{sec:analysis}

\subsection{Stellar atmospheric parameters}

\begin{table}
\caption{Stellar atmospheric parameters, orbital solution, and physical properties of the \tic\ system. Symmetric errors are reported with $\pm$ and asymmetric errors are reported in brackets and correspond to the difference between the median and the 16$^{th}$ (lower value) and 84$^{th}$ (upper value) percentile.}              
\label{tab:parameters}      
\centering   
\begin{tabular}{l c}          
\hline\hline                        
Parameter & value\\
\hline 
Gaia \\
Source ID & 6391016653342250240  \\
BP & $10.865563 \pm 0.01$ \\
RP & $10.209588 \pm 0.01$ \\
Parallax [mas] & $2.5025 \pm 0.0236$ \\ \\

Spectroscopy \\
$\rm T_{\rm eff}$ $\rm(K)$
& $6280 \pm 85$ \\

$\log g$ (dex)  
& $4.01 \pm 0.13$ \\

$\xi_{\rm t}\, (\rm km\,s^{-1})$
& $1.17 \pm 1.50$ \\

$v_{\rm mac}\, (\rm km\,s^{-1})$
& $4.67 \pm 1.50$ \\

Vsin$i$ (km\,s$^{-1}$)
& $ \leq 2$ \\

$\rm [Fe/H]$ & $0.20 \pm 0.06$ \\ \\

Orbital solution \\
$\rm T_{\rm 0}$ [JD] &  $2458337.730695_{(1264)}^{(1701)}$\\
Period [d] & $38.195178_{(121)}^{(58)}$ \\
$R_1 / a$ & $0.0164_{(5)}^{(2)}$\\
$R_2 / R_1$ & $0.0789_(5)^{(10)}$ \\
$b$ & $0.034_{(10)}^{(89)}$ \\
$\rm h_{\rm 1,WASP}$ & $0.7387_{(7)}^{(12)}$ \\
$\rm h_{\rm 2,WASP}$ & $0.2000_{(6)}^{(8)}$ \\
$\rm h_{\rm 1,TESS}$ & $0.8267_{(18)}^{(2)}$  \\
$\rm h_{\rm 2,TESS}$ & $0.2023_{(80)}^{(100)}$ \\
$\rm h_{\rm 1,NGTS}$ & $0.8214_{(18)}^{(1)}$  \\
$\rm h_{\rm 2,NGTS}$ & $0.2042_{(8)}^{(7)}$\\
$\sigma_{\rm WASP}$  & $0.00756_{(34)}^{(2)}$ \\
$\sigma_{\rm TESS}$  & $0.00045_{(6)}^{(16)}$ \\
$\sigma_{\rm NGTS}$  & $0.00804_{(22)}^{(1)}$ \\
$K_1$ [km\,s$^{-1}$] & $7.013_{(36)}^{(49)}$ \\
$f_s$ & $-0.038_{(22)}^{(9)}$ \\
$f_c$ & $0.545_{(3)}^{(1)}$ \\
$V_0$ [km\,s$^{-1}$] & $-12.14_{(60)}^{(80)}$ \\
$dV_0 / dt$ [km\,s$^{-1}$\,d$^{-1}$] & $-0.003_{(2)}^{(1)}$ \\
$J$ [km\,s$^{-1}$] & $0.054_{(2)}^{(102)}$ \\ \\

Physical properties \\
$M_1$ [$M_{\odot}$] & $1.514_{(37)}^{(37)}$ \\
$R_1$ [$R_{\odot}$] & $2.159_{(37)}^{(37)}$ \\
$M_2$ [$M_{\odot}$] & $0.148_{(3)}^{(3)}$ \\
$R_2$ [$R_{\odot}$] & $0.171_{(3)}^{(3)}$ \\
Age [Gyr] & $9.3_{(1)}^{(1)}$ \\

\hline
\end{tabular}
\end{table}

 We used wavelet analysis to extract atmospheric parameters from the co-added eight CORALIE spectroscopic observations of \tic\ (Sect.\,\ref{sec:spec}) using the methodology set out in \citet{2018A&A...612A.111G,2019A&A...626A.119G}.
 The co-added spectrum was re-sampled between 450-650\,nm with $2^{17}$ values. The wavelet coefficients $W_{i=4-14, k}$ were calculated \citep[see Fig.\,2 of][]{2018A&A...612A.111G}
 and fitted against the same coefficients from model spectra in a Bayesian framework. We initiated 100 walkers and generated 100,000 draws as a burn-in phase. We generated a further 100,000 draws to sample the PPD for $T_{\rm eff}$, [Fe/H], $V\sin\,i$ and $\log g$. 
 The wavelet method for CORALIE spectra can determine $T_{\rm eff}$ to a precision of $85$\,K, [Fe/H] to a precision of 0.06\,dex and $V \sin i$ to a precision of 1.35\,km\,s$^{-1}$.
 \tic\ has a projected rotation below 0.5\,km\,s$^{-1}$ and so we do not attribute any uncertainty to our measurement of $V \sin i$. Measurements of $\log g$ from wavelet analysis are not reliable beyond confirming dwarf-like gravity ($\log g \approx 4.5$ dex). Subsequently, we fit the wings of the magnesium triplets with spectral synthesis by fixing $T_{\rm eff}$, [Fe/H] and $V\sin i$ and changing $\log g$ until an acceptable fit was found.  All our derived parameters for \tic\ are set out in full in Table \ref{tab:parameters}.

\subsection{Global modelling}\label{Sect:joint}

We collectively modelled \wasp, \tess\, and NGTS photometry with the \textit{CORALIE} radial velocity measurements.  Preliminary modelling of each photometric dataset found consistent transit depths (to within 1-$\sigma$) so we decided to fit a common transit depth ($k=R_2/R_1$). Our model used the method described by \citet{2016A&A...591A.111M} to solve Kepler's equations and the analytical approximation presented by \citet{2019A&A...622A..33M} to describe an object eclipsing a star with limb-darkening described by the power-2 law. We fitted the decorrelated limb-darkening parameters $h_1$ \& $h_2$ from Eqn. 1 \& 2 of \citet{2018A&A...616A..39M}. Following the suggestion by \citet{2018A&A...616A..39M}, Gaussian priors were centred on interpolated values of $h_1$ and $h_2$ (from Tab.\,2 of \citet{2018A&A...616A..39M} via the \textsc{pycheops} python package)
with widths of 0.003 and 0.046 respectively. The similarity between \textit{NGTS} and \textit{TESS} transmission filters is such that they could be fitted with common limb-darkening priors. We used a different limb-darkening prior for the WASP  filter which is  bluer than \textit{TESS} and \textit{NGTS}. 

\begin{figure}
    \centering
    \includegraphics[scale=0.66]{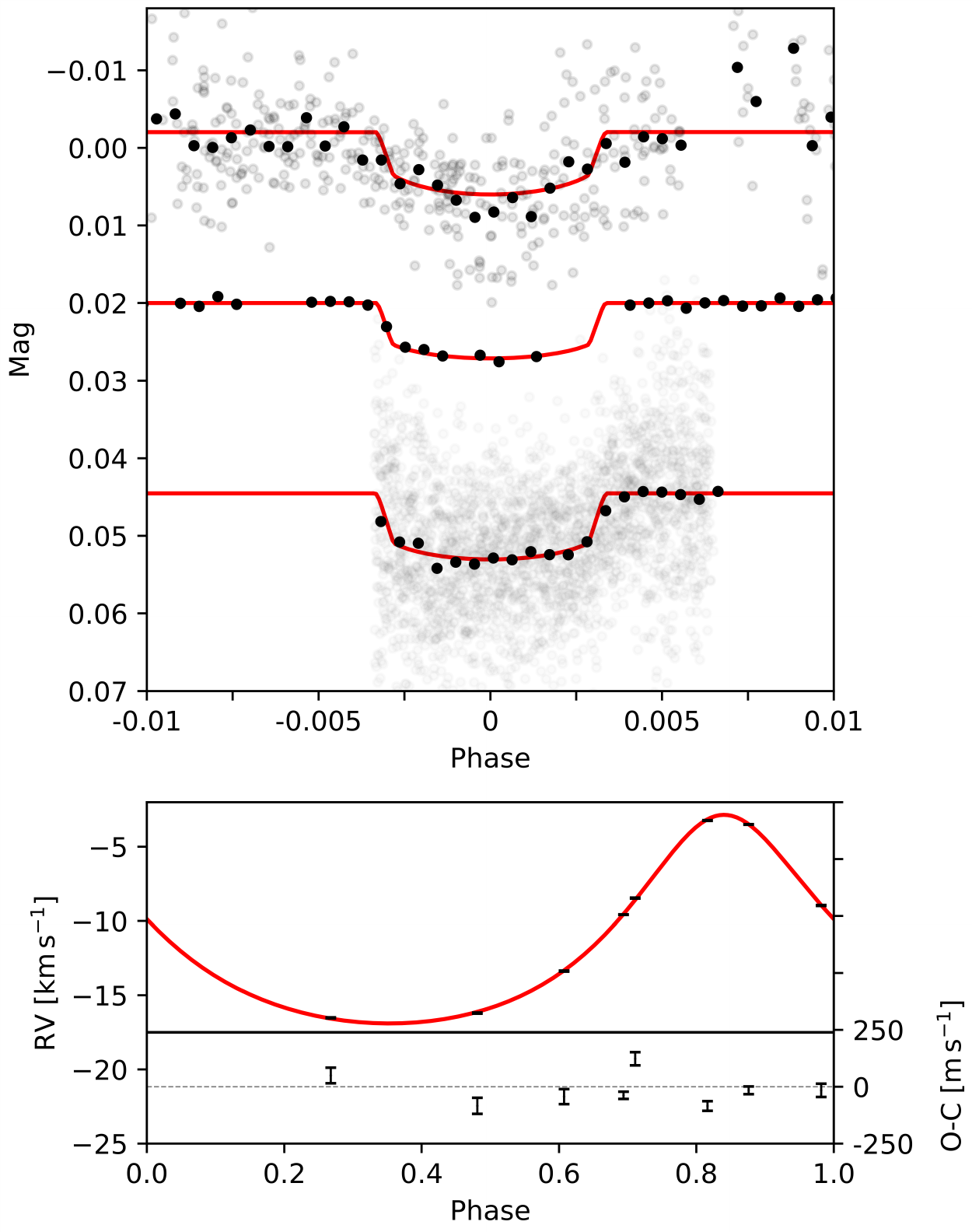}
    \caption{Orbital solution for TIC-238855958. Transit photometry from \textit{WASP} (top), \textit{TESS} (upper middle), and \textit{NGTS} (lower middle) with corresponding best-fitting models (red). The darker points show all three light curves at 30 \,min cadence. 
    The bottom panel shows CORALIE radial velocity measurements (black) with best-fitting model (red). }
    \label{fig:Figure_d}
\end{figure}

Our model vector included $T_0$, the orbital period, $P$, $R_1 / a$, $k=R_2/R_1$, $b$, independent values of the photometric zero-point, $zp$, $h_1$ and $h_2$ for each bandpass, the semi-amplitude, $K_1$, the radial velocity zero- of the primary star, $\gamma$, and the change radial velocity of the primary star with time, $d(\gamma)/dt$. Instead of fitting the argument of the periastron ($\omega$) and the eccentricity ($e$), we used  $f_c = \sqrt{e} \cos \omega$ and  $f_s = \sqrt{e} \sin \omega$ since these have a uniform prior probability distribution and are not strongly correlated with each other. We also include a jitter term  ($J$) to account for spot activity which can introduce noise in to the radial velocity measurements \citep{2006ApJ...642..505F}. This was added in quadrature to the uncertainties associated with each RV measurement. We fit a similar term for each photometric data set, $\sigma$, which was also added in quadrature.
We sample parameter space using the Bayesian sampler described in Sect.\,\ref{sec:precovery}. We ran 50 Markov chains for 100\,000 draws and discarded the first 50\,000 as a burn-in phase - visual checks ensured convergence was achieved well before the 50,000$^{th}$ draw. We selected the trial step with the highest value of log-likelihood as the measurement for each parameter. Asymmetric uncertainties were calculated using the differences between the measurement and the $16^{\rm th}$ and $84^{\rm th}$ percentiles of the cumulative posterior probability distribution (PPD). Fitted parameters are reported in Tab.\,\ref{tab:parameters} and shown in Fig.\,\ref{fig:Figure_d}.

\subsection{Physical properties}

We used the \textsc{isochrones} python package
\citep{2015ascl.soft03010M} to measure the physical properties of the host star. Our vector of model parameters included the Gaia magnitudes $BP$ and $GP$ and parallax with Gaussian priors centred on values reported from GAIA DR2 \citep{2018A&A...616A...1G} with widths of 0.01 magnitudes, $\rm T_{\rm eff}$, $\log g$, and [Fe/H] with prior centres and width equivalent to values and errors reported in Table \ref{tab:parameters} respectively. We use \textsc{emcee} to sample the posterior distributions of each parameters. We discarded 10\,000 draws as a burn-in phase before drawing the sample number of post burn-in draws used in Sect. \ref{Sect:joint}. The PPD for the physical parameters associated with the fit ($M_1$, $R_1$, age) were used to calculate the physical properties of the M-dwarf. We calculated the PPD for $R_2$ by multiplying the PPDs for $k$ and $R_1$. Calculating $M_2$ required solving the spectroscopic mass function,
\begin{equation}\label{eqn:mass_func}
    f(M) = \frac{\left( M_2 \sin i\right) ^3}{\left( M_\star + M_2\right)^2} = \left( 1 - e^2 \right)^{3/2} \frac{PK_1^3}{2 \pi G},
\end{equation}
where $G$ is the gravitational constant. For each step in the respective samplers, we evaluated the left hand side of Eqn. \ref{eqn:mass_func} and solved for $M_2$ using the corresponding value of $M_\star$. We assumed both stars are coeval. 

\section{Discussion and conclusions}\label{sec:conclusion}

\begin{figure}
    \centering
    \includegraphics[scale=0.6]{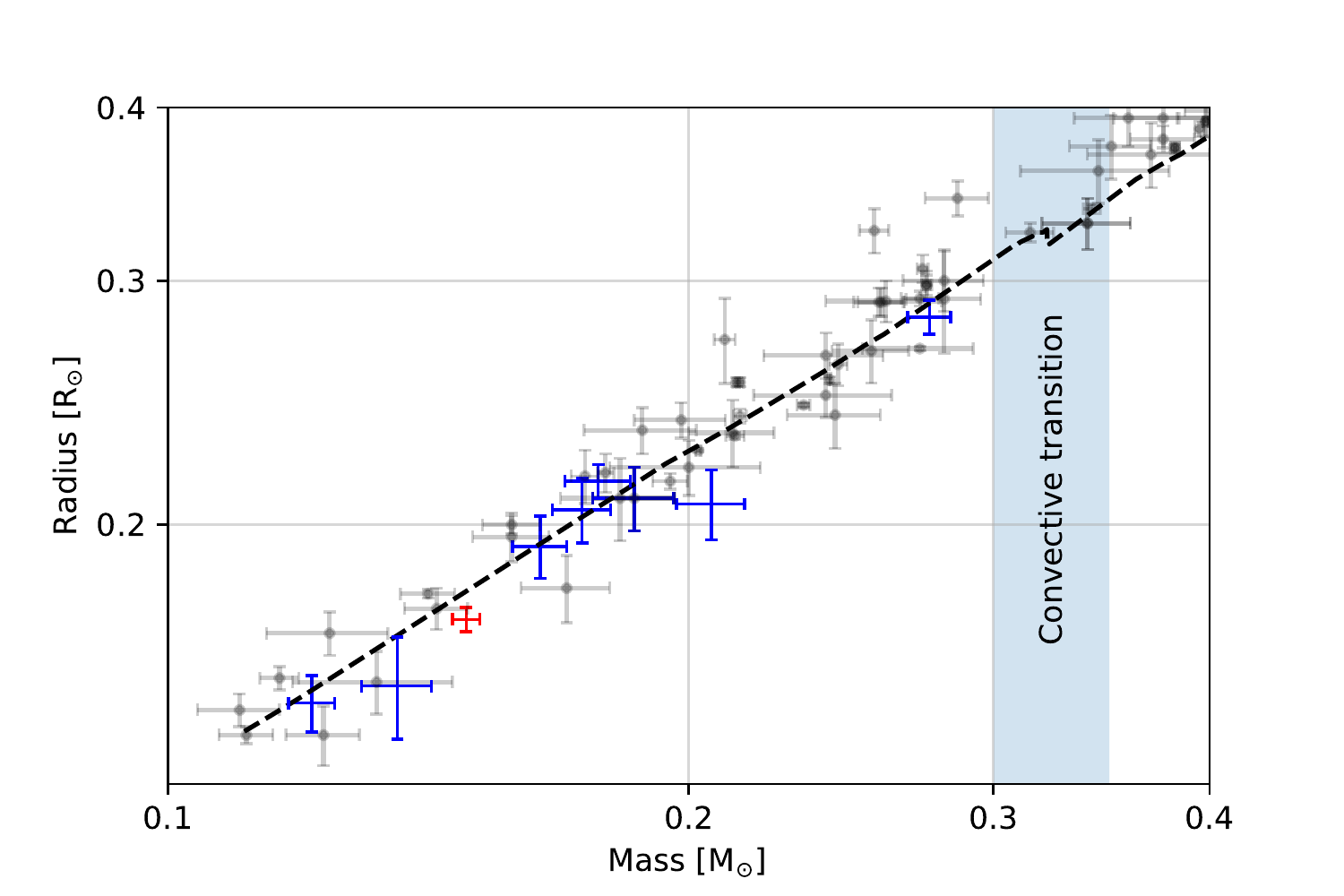}
    \caption{Mass-radius diagram for late M-dwarfs in eclipsing systems. We plot the M-dwarf companion of TIC-238855958 in red, M-dwarfs measured within the EBLM project in blue, and M-dwarfs 
     with masses and radii known to better than 10\%
    \citep[from Table 4 of][and references therein]{2018AJ....156...27C} 
    in black. We overplot the best-fitting isochrone for TIC-238855958 (black-dashed). }
    \label{fig:mdr}
\end{figure}

Our global modelling shows \tic\ to be an eccentric, long period (38.2\,days), F-M eclipsing binary.  The primary F-star appears to have turned-off the main-sequence, but has made little progress through the red-giant branch phase.  Due to its mass (M$_1$=$1.514 \pm 0.037\,M_\odot$), the primary star is set to transition through the post-main sequence blue hook that marks the passage from central hydrogen burning to shell burning. The M-dwarf has a mass of M$_2$=$0.148 \pm 0.003\,M_\odot$ and a radius of R$_2$=$0.171 \pm 0.003\,R_\odot$, making it one of the best characterised low-mass stars in terms of its mass and radius.  This is shown in Fig.\,\ref{fig:mdr} where we compare the mass and radius of the M-dwarf 
with other well measured low mass stars.  We find it to be 2.3-$\sigma$ smaller than expected compared with evolutionary models, see Fig.\,\ref{fig:mdr}. This is consistent with J2308$-$46 and J1847$+$39, which are deflated by at least 1-$\sigma$ \citep{2019A&A...626A.119G}. In general, M-dwarfs appear to be systematically inflated and cooler than predicted from models of stellar evolution \citep[see Fig.\,9 of][]{2017ApJ...844..134L}. It would be interesting to measure the secondary eclipse depth of the M-dwarf during the \tess\ extended mission, in order to compare its luminosity with evolutionary models. 

\tic\  is the first success of the NGTS Mono-transit Working Group and its discovery paves the way to characterising long-period eclipsing systems. This work highlights the pivotal role archival photometric databases play in the recovery of orbital periods in conjunction with current photometric and spectroscopic instruments. Although this system is a long-period eclipsing binary, the transit depth is consistent with a giant planet around a solar-type star and so our method will be just as efficient at finding long-period transiting exoplanets.

\section*{Acknowledgements}

The NGTS facility is operated by the consortium institutes with support from the UK Science and Technology Facilities Council (STFC) under projects ST/M001962/1 and ST/S002642/1. 
Contributions at the University of Geneva by FB, LN, ML, OT, and SU were carried out within the framework of the National Centre for Competence in Research "PlanetS" supported by the Swiss National Science Foundation (SNSF).
The contributions at the University of Warwick by PJW, RGW, DLP, DJA, DRA, SG, and TL have been supported by STFC through consolidated grants ST/L000733/1 and ST/P000495/1. DJA acknowledges support from the STFC via an Ernest Rutherford Fellowship (ST/R00384X/1).
The contributions at the University of Leicester by MGW and MRB have been supported by STFC through consolidated grant ST/N000757/1.  SLC acknowledges support from the STFC via an Ernest Rutherford Fellowship (ST/R003726/1)
JSJ is supported by funding from Fondecyt through grant 1161218 and partial support from CATA-Basal (PB06, Conicyt).
ACC acknowledges support from
the Science and Technology Facilities Council (STFC) consolidated
grant number ST/R000824/1.
MNG acknowledges support from the Juan Carlos Torres Fellowship.

\bibliographystyle{mnras}
\bibliography{references} 

\label{lastpage}
\end{document}